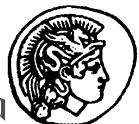

# Quantum Hall stripe phases at integer filling factors

E. Demler[a], D.-W. Wang[b],*, S. Das Sarma[b], B.I. Halperin[a]

[a]*Department of Physics, Harvard University, Cambridge, MA 02138, USA*
[b]*Condensed Matter Theory Center, Department of Physics, University of Maryland, College Park, MD 20742, USA*

## Abstract

We consider stripe formation in quantum Hall (QH) systems at integer filling factors. We use Hartree–Fock calculations to obtain the phase diagram of bilayer QH systems at $\nu = 4N + 1$ in a tilted magnetic field. We derive and analyze an effective low energy theory for the stripe phases, which may be present in such systems. We discuss the possibility of stripe formation in wide well systems in a tilted magnetic field and suggest that the resistance anisotropy, observed recently by Pan et al. [Phys. Rev. B 64 (2001) 121305], may be due to the existence of a skyrmion stripe phase. © 2002 Elsevier Science Ltd. All rights reserved.

## 1. Introduction

Macroscopic degeneracy of electrons in Landau levels amplifies the importance of Coulomb interaction for the quantum Hall (QH) systems and gives rise to a variety of unusual phenomena. The most famous example is the existence of the fractional QH effect, where at certain filling factors the system becomes incompressible, i.e. it acquires gaps for making charge excitations [1]. Another manifestation of Coulomb interaction is the formation of stripe phases in high Landau levels at half-integer filling factors [2–8]. Recently isospin stripes at integer filling factors have been proposed theoretically by Brey and Fertig for bilayer systems at $\nu = 4N + 1$ [9], where the two layers are labeled by an isospin index, $\sigma$.

In this paper we extend the ideas of [9] to the case of tilted magnetic fields and suggest the possibility of two additional stripe phases: an isospin stripe phase (ISt) with winding and a skyrmion stripe phase. We derive an effective low energy theory for these stripe phases and study its consequences. We show that this construction has a natural generalization to two-dimensional electron gases in wide wells in tilted magnetic fields and argue that these phases may be relevant for understanding the resistance anisotropy recently reported in Ref. [10].

## 2. Hartree–Fock calculations for bilayer systems at $\nu = 4N + 1$

Properties of electrons in bilayer QH systems at $\nu = 4N + 1$ have been a subject of active theoretical and experimental research (see Refs. [11–17], and references therein). In a perpendicular magnetic field and small $d/l_0$ (here $d$ is the distance between the layers and $l_0 = (c/eB_\perp)^{1/2}$ is the magnetic length) it is commonly accepted that the ground state is a Halperin (1,1,1) state [18,19]. When tunneling between the layers is small, such a state corresponds to spontaneous interlayer coherence with a Goldstone mode [20,21]. Brey and Fertig suggested that for $N > 0$ as $d$ is increased the interlayer coherent phase is unstable to the formation of isospin stripes, i.e. oscillations in the charge distribution between the layers with the total charge density in the two layers fixed. The origin of such isospin stripe order is a competition between the exchange and the direct Coulomb interaction. Exchange favors accumulating all the electrons in one layer (in order to maximize the isospin exchange field), whereas direct

* Corresponding author. Tel.: +1-301-405-6173; fax: +1-301-314-9465.
*E-mail address:* weiwang@glue.umd.edu (D.W. Wang).

Coulomb energy is lower when the electron density is distributed uniformly between the layers.

It may be worthwhile to recapitulate the coherence underlying the Halperin (1,1,1) phase [18] in the context of the bilayer QH system. The two-component generalization, the Halperin $(m_1, m_2, n)$ state, of the Laughlin wavefunction proposed by Halperin in 1983 has the form

$$\Phi_{m_1,m_2,n}[z] \equiv \prod_{i<j\leq N_\uparrow} (z_i - z_j)^{m_1} \prod_{k<l\leq N_\downarrow} (z_{[k]} - z_{[l]})^{m_2} \prod_{a=1}^{N_\uparrow} \prod_{b=1}^{N_\downarrow}$$
$$\times (z_a - z_{[b]})^n \prod_{c=1}^{N} \exp\left(\frac{-|z_c|^2}{4}\right), \quad (1)$$

where $z_j = (x_j + iy_j)/l_0$, is the standard two-dimensional layer coordinate of the $j$th electron, $m_1/m_2$ are both odd integers (to preserve the Pauli exclusion principle), $[j] \equiv N_\uparrow + j$, and $N_\uparrow/N_\downarrow$ are the number of electrons in the two components ($N = N_\uparrow + N_\downarrow$) [12]. For bilayer systems one could think of $N_\uparrow/N_\downarrow$ as the electron density in the isospin state associated with the layer index. The special case, $m_1 = m_2 = n = 1$, of the above wavefunction describes a coherent bilayer state with the total filling factor of unity with a fixed total density, $N$, but with indefinite number of electrons in each layer. The Halperin (1,1,1) state has been extensively studied in the literature [11,12], and is an example of a class of novel many-body ground states for bilayer systems which exhibit spontaneous interlayer coherence [14–18].

If the magnetic field is tilted away from perpendicular to the sample, so that there is a nonzero in-plane field $B_\parallel$, a variety of new phases may occur, which we investigate in this paper. We assume that the first $4N$ Landau levels are completely full, with both spin states and both layers populated, and that the remaining electrons, of which there are one per flux quantum, are completely spin polarized, but distributed between the two layers (i.e. isospin unpolarized). We introduce an isospin index $\sigma = \uparrow, \downarrow$ to distinguish the layers, and define isospin density operators $I_z = (\Psi_\uparrow^\dagger \Psi_\uparrow - \Psi_\downarrow^\dagger \Psi_\downarrow)/2$ and $I_+ = \Psi_\uparrow^\dagger \Psi_\downarrow$, where $\Psi_\sigma$ is the annihilation operator for an electron in layer $\sigma$ at a given point in the $x$–$y$ plane. We also use the Landau gauge, $\vec{A}(\vec{r}) = (0, B_\perp x, B_x y - B_y x)$, so that single electron states are $\phi_{nk\sigma}(\vec{r}) = L_y^{-1/2} \delta(z + \sigma d) H_n(x + kl_0^2) e^{iky}$, where $k$ is the wavevector that labels states within the Landau level $n$ ($k = 2\pi m/L_y$, and $m$ is an integer that goes from 1 to the Landau level degeneracy $N_\phi$), $L_y$ the length of the system in the $y$ direction, and $\sigma = \pm 1/2$.

For $B_\parallel = 0$, the coherent (1,1,1) phase has $\langle I_+ \rangle$ equal to a real, nonzero constant, and in the absence of stripes, $\langle I_z \rangle$ is also a constant, independent of position. (If the two layers are symmetric, and there is no spontaneously broken symmetry, we then have $\langle I_z \rangle = 0$). In the stripe phase of Ref. [9], $\langle I_z \rangle$ varies periodically in space along one direction (which we take here to be the $x$-direction). The magnitude of $\langle I_+ \rangle$ will also vary periodically, but its phase remains constant, so that $\langle I_+ \rangle$ is real.

For nonzero $B_\parallel$, in the absence of stripes, we may distinguish two possible quantized Hall states. In one, which is often denoted [11–13] the 'commensurate phase', but which we call here the 'isospin spiral phase' (ISp), $\langle I_+ \rangle$ has a form $I_0 e^{i\vec{P}\cdot\vec{r}}$, which has constant magnitude and a phase which varies linearly in space, with $\vec{P} = 2\pi d\vec{B}_\parallel \times \hat{z}/\Phi_0$ and $\Phi_0 = hc/e$. In the other 'incommensurate state', the phase of $\langle I_+ \rangle$ varies more slowly in space (see, e.g. [12,13,22]). In the Hartree–Fock (HF) calculations reported in this paper, we ignore entirely the phase variation in the incommensurate state, and take it to be constant in space; and we denote this as the 'coherent phase'. This greatly simplifies the calculations, and should introduce only a small error in the energy of the state, except very close to the commensurate–incommensurate phase boundary [12,13].

For nonzero $B_\parallel$, if stripes are present, the energy will depend on the orientation of the stripes relative to the in-plane field. In experiments the direction of the parallel magnetic field is fixed, and the stripe direction is presumably determined by the condition of minimizing the ground state energy. Here, we fix the direction of the stripe order and adjust the direction of the parallel magnetic field. We will consider two cases, where $B_\parallel$ is either perpendicular to the stripes ($\vec{B}_\parallel \| \hat{x}$) in our notation, or parallel to them ($\vec{B}_\parallel \| \hat{y}$). In either case, stripe order in $\langle I_z \rangle$ may coexist with commensurate spiral order or with incommensurate order in $\langle I_+ \rangle$. We denote the commensurate phase with stripes parallel to $B_\parallel$ as an 'isospin spiral stripe phase' (ISpSt), while we designate the commensurate phase with stripes perpendicular to $B_\parallel$ as 'isospin skyrmion stripe phase' (ISkSt) (see Fig. 1(b) and (c)). The latter phase has finite topological isospin density $\rho_{\text{topo}} = \epsilon^{abc} \epsilon_{\alpha\beta} I^a \partial_\alpha I^b \partial_\beta I^c$ and therefore carries a charge density wave in addition to the isospin density wave [23]. The ISt without spiral order (see Fig. 1(a)) will be denoted simply as an ISt. This phase was called 'unidirectional coherent charge density wave state' in Ref. [9].

In Fig. 2, we show the phase diagram obtained from a $T = 0$ HF calculation at $\nu = 5$, in a case where $B_\parallel$ is assumed parallel to the stripes. Phase I is the incommensurate coherent phase present for small values of $d/l_0$ and small interlayer tunneling $t$. In this phase the interlayer exchange interaction is strong enough to keep the densities in each layer uniform and prevent the relative phase between the layers from winding at wavevector $\vec{P}$. As the tunneling is increased it becomes energetically favorable to have a phase winding at wavevector $\vec{P}$ and we find a commensurate ISp phase without stripes (phase II on Fig. 2). When $d/l_0$ is increased and $t$ is kept small, we find the striped ISt phases (phase III on Fig. 2). It is useful to point out that transition between the coherent and ISt phase is continuous, so the stripe order develops gradually in the ISt phase with a simultaneous suppression of the interlayer coherence. When both $d/l_0$ and $t$ are increased we find the ISpSt phase with both stripes and phase winding (phase IV on Fig. 2). Another regime where we find the ISpSt phase is small $d/l_0$ and large $t$ (phase V on Fig. 2). This phase is related to the pseudospin canted phase proposed in Ref. [24]. The latter paper proposed that for short range interactions a parallel magnetic field may induce spontaneous imbalance in the charge density between the two layers, and adding a long range interaction will lead to the appearance of large domains. This is indeed what we find: long range Coulomb interaction prevents the appearance of real charge imbalance between the layers, and the system goes into a spiral stripe phase with very long period $a$ of $I_z$ modulation ($a \gg l_0$). HF calculations for phase V have also been reported in Ref. [25]. HF calculations predict qualitatively similar phase diagrams for all filling factors $\nu = 4N + 1$, including $\nu = 1$. We note, however, that this approach does not include the possibility of two decoupled $\nu = 1/2$ states in the two layers (on top of the filled $4N$ levels), which is more favorable for smaller $n$. We therefore expect that the phases discussed in this paper are more likely to be found for $N > 0$. Similar states can occur at filling factors $4N + 3$, interchanging the role of holes and electrons in the highest Landau level.

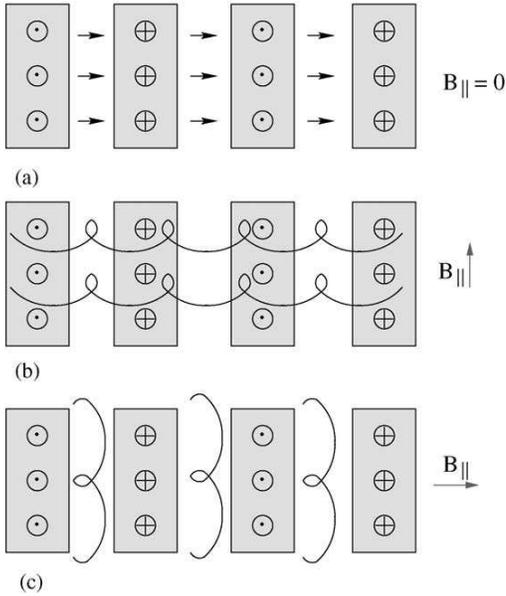

Fig. 1. ISts discussed in this paper. Shaded areas show isospin up (⊙) and down (⊕) domains, horizontal arrows and spirals indicate transverse isospin $(I_x, I_y)$ order parameter, and long arrows in the right hand side show the direction of $B_\parallel$. (a) Is the ISt (this corresponds to phase III on Fig. 2 in the limit $B_\parallel = 0$). (b) Is the ISpSt (this corresponds to phases IV and V on Fig. 2). (c) Shows the ISkSt. We propose that the resistance anisotropy observed in wide well experiments in Ref. [10] is due to formation of a spin skyrmion stripe phase. The latter is similar to ISkSt (c), but involves the actual spin of the electrons, rather than the isospin.

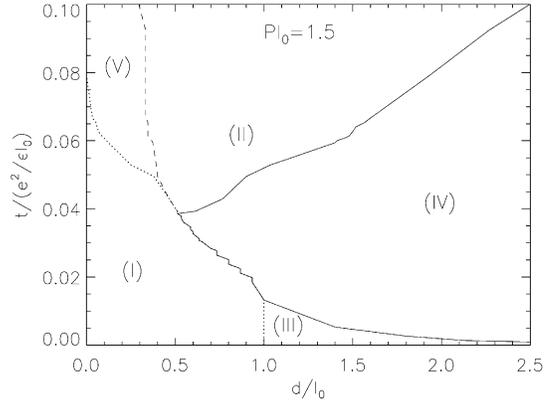

Fig. 2. Phase diagram for bilayer system at $\nu = 5$ in the presence of parallel magnetic field $B_\parallel dl_0/\Phi_0 = 1.5$. Phase I is an incommensurate interlayer coherent phase, phase II is an ISp, phase III is an ISt, phases IV and V are ISpSts. ($t$ and $d$ refer to tunneling amplitude and layer separation, respectively).

To provide a better picture of these phases let us consider a Slater determinant state in bilayer systems at $\nu = 4N + 1$

$$\prod_k \left( e^{i\alpha_k} \cos\frac{w_k}{2} c^\dagger_{k-Q_y/2\downarrow} + e^{-i\alpha_k} \sin\frac{w_k}{2} c^\dagger_{k+Q_y/2\uparrow} \right) |0\rangle, \quad (2)$$

where $\alpha_k = kQ_x l_0^2$, $c^\dagger_{k\sigma}$ creates an electron in state $\phi_{Nk\sigma}(\vec{r})$, and $|0\rangle$ denotes a state of $4N$ filled Landau levels in both layers and both spin components. When $w_k$ is constant, the wavefunction (2) describes nonstripe phases: fully isospin polarized uniform phases for $\vec{Q} = 0$ ($w = 0$ and $\pi$ correspond to all electrons in the left and the right layers, respectively), and a spiral phase for finite $\vec{Q}$. When $w_k$ changes periodically with $k$, we obtain stripe phases. The case $\vec{Q} = 0$ has been considered by Brey and Fertig in Ref. [9] and corresponds to phase III in Fig. 2. The wavefunction (2) is written in such a way that it has oscillations in $I_z$ along the $x$ axis. When $\vec{Q} = (P, 0)$ (this requires $\vec{B}_\parallel \| \hat{y}$) we have a spiral stripe phase, and when $\vec{Q} = (0, P)$ (such winding requires $\vec{B}_\parallel \| \hat{x}$) we obtain a skyrmion stripe phase. For layers of zero width and within HF approximation we find that the ISpSt phase is always lower in energy than the ISkSt phase. However, the energies of the two states are very close, so when the finite layer width is taken into account, the ISkSt phase may be favored energetically [26].

## 3. Effective theory for the stripe phases in bilayer systems at $\nu = 4N + 1$

To derive an effective low energy theory for the spiral stripe phase shown in Fig. 1(b) it is convenient to start with an ISt in which $w_k$ take values 0 or $\pi$ only. This phase has no interlayer coherence, and domain boundaries between ↑ and ↓ are sharp with two counter-propagating edge states of different isospins (see

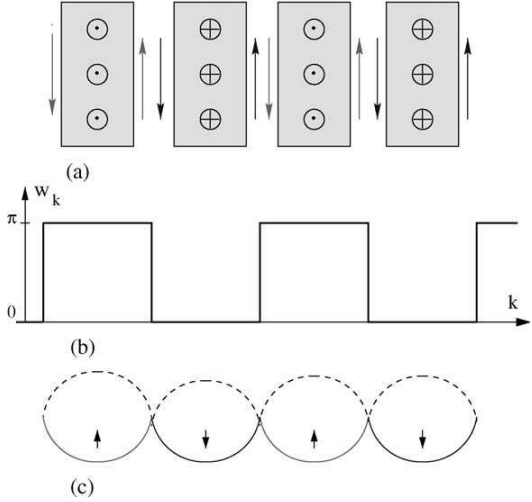

Fig. 3. (a) Shows an ISt without interlayer coherence. On (b) we show that it is described by wavefunction (2) with $w_k$ taking values 0 and $\pi$ only. (c) Gives the HF potential for the isospin up and down electrons and indicates the existence of the isospin ↑ and ↓ edge states on domain walls. These edge states are shown as upward and downward arrows on (a).

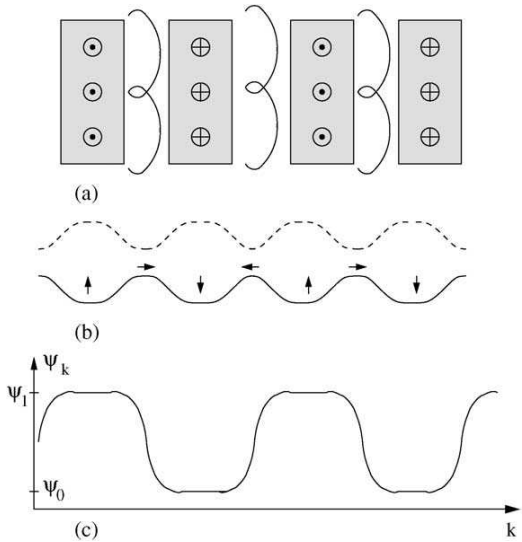

Fig. 4. ISkSt with oscillations in $I^z$ and $I^x$, $I^y$. (a) Shows up and down domain separated by regions of the spiral phase. (b) Shows that this state may be thought of as coming from hybridization of the internal edge states (see Fig. 3) at finite wavevector $q_y$, that opens a gap in the quasiparticle spectrum. Appropriate $\psi_k$ are shown in (c). Here $I^x$ and $I^y$ wind at some wavevector perpendicular to the $I^z$ modulation, so there is a topological spin density that results in charge modulation.

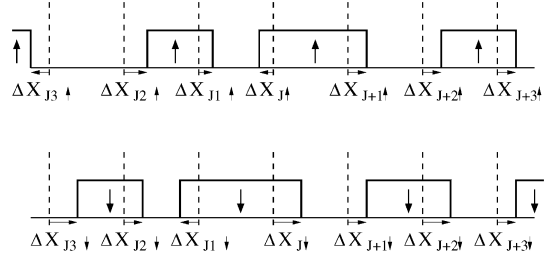

Fig. 5. Distorted isospin stripe from Fig. 3. Note that a displacement of domain edges in the positive $x$ direction leads to an extra density on the right hand side of each of the domains and to a missing density on the left side. Hence, $\Delta\rho_{j\uparrow} = (-)^j \Delta X_{j\uparrow}$ and $\Delta\rho_{j\downarrow} = (-)^{j+1} \Delta X_{j\downarrow}$.

Fig. 3). Various patterns of the edge states hybridization (which open a gap in the quasiparticle spectrum) will allow us to connect this phase to all the other stripe phases (see, e.g. Fig. 4). Following our HF calculations we start from a configuration where stripes are parallel to $B_\parallel$ and are in the $y$ direction. On the edge $j$, electron operators may be bosonized using $\psi_{j\sigma} = e^{i\phi_{j\sigma}}$ (see Ref. [27] and references therein). Fermion operators of a given isospin change their chirality between the neighboring stripes (see Figs. 3 and 5), so the displacements and resulting density changes at the isospin up and down edges are $\Delta X_{j\uparrow} = (-)^j \Delta\rho_{j\uparrow} = \partial_y \phi_{j\uparrow}$ and $\Delta X_{j\downarrow} = (-)^{j+1} \Delta\rho_{j\downarrow} = \partial_y \phi_{j\downarrow}$ (from now on all lengths are measured in the units of $l_0$). Following [8,28–31] it is convenient to separate fluctuations of the edge position $u_j = (\Delta X_{j\uparrow} + \Delta X_{j\downarrow})/2 = (-)^j(\Delta\rho_{j\uparrow} - \Delta\rho_{j\downarrow})/2 = \partial_y \phi_{j+}$ and fluctuations in the domain wall width $n_j = (\Delta X_{j\uparrow} - \Delta X_{j\downarrow})/2 = (-)^j(\Delta\rho_{j\uparrow} + \Delta\rho_{j\downarrow})/2 = \partial_y \phi_{j-}$ (they may be thought of as charge fluctuations, since shifting the edge positions relative to each other creates a region of lower or higher electron density), where we introduced $\phi_{j\pm} = (\phi_{j\uparrow} \pm \phi_{j\downarrow})/2$. The relative phase between the up and down electrons $2\phi_{j-}$ is what corresponds to the phase of the transverse isospin $I_+$. The staggered relation between the displacements and the densities makes it convenient to separate the staggered and uniform components of every field $\phi_{j+} = (-)^j \phi_j^s + \phi_j^o$, $\phi_{j-} = (-)^j \theta_j^s + \theta_j^o$, $u_j = (-)^j u_j^s + u_j^o$, and $n_j = (-)^j n_j^s + n_j^o$; with $u_j^s = \partial_y \phi_j^s$, $u_j^o = \partial_y \phi_j^o$, $n_j^s = \partial_y \theta_j^s$, and $n_j^o = \partial_y \theta_j^o$. Simple arguments can now be used to write an effective Hamiltonian for the ISt

$$\mathcal{H}_{\text{eff}} = \int_{\vec{r}} \left\{ \frac{m_1^2}{2}(u^s)^2 + \frac{1}{2}(K_x^\prime (\nabla_x u^o)^2 + K_y (\nabla_y^2 u^o)^2) \right.$$

$$+ \frac{1}{2}(C_x (\nabla_x \theta^o)^2 + C_y (\nabla_y \theta^o)^2) + \frac{m_2^2}{2}(\theta^s)^2$$

$$\left. - t\cos(\theta^o - Px) + D(\nabla_x \theta^o)^2(\nabla_y u^o)^2 + \Gamma_0 u^s \nabla_x u^o \right\} \quad (3)$$

Here $\int_{\vec{r}} = \int dx\,dy$, $P = |\vec{P}| = 2\pi B_\parallel d/\Phi_0$. The first term in Eq. (3) comes from the fact that $u^s$ is a massive variable that changes the width of the spin up domains relative to the spin down domains, i.e. isospin polarization ($I^z$) of the system. By contrast, $u^o$, shifts up and down domains equally and the second and third in Eq. (3) describe soft stripe position fluctuations. In the absence of parallel magnetic field the direction of stripes is arbitrary, so the gradient energy for $u^o$ is of the smectic type [8]. Parallel magnetic field selects a particular direction of the stripe orientation and gives finite stiffness to stripe fluctuations in the $y$ direction as will be discussed later. Exchange interactions penalize gradients of the relative phase between the up and down electrons but do not select the phase itself. So in the absence of tunneling, fluctuations in $\theta^o$ do not change the energy of the system in the long-wavelength limit, but the fluctuations of $\theta^s$ are massive, as described by Eq. (3). The tunneling favors a phase $\theta^o$ that winds in the $x$ direction at wavevector $P$ (assuming $\vec{B}_\parallel \| \hat{y}$) and leads to the seventh in Eq. (3). Twisting of $\theta^o$ is present for any value of the parallel magnetic field and, according to our HF calculations, stabilizes the direction of the stripes that is perpendicular to the gradient of $\theta^o$. The particular form of the eighth term in Eq. (3) is fixed by the observation that reversing the direction of the parallel magnetic field will not change orientation of the stripes. Finally, the last term of Eq. (3) takes into account that a change in polarization $u^s$ will lead to a change in the stripe periodicity $\nabla_x u^o$.

Hamiltonian (3) must be supplemented by the commutation relations between the internal edge state densities $[\rho_{j\alpha}(q_y), \rho_{j'\beta}(q'_y)] = \delta_{jj'}\delta_{\alpha\beta}q_y\delta(q_y + q'_y)$. They give rise to the terms in the Euclidean space-time action $S_\tau = i\int d\tau\,dy \sum_{j\sigma}(-)^j \sigma \partial_\tau \phi_{j\sigma}\partial_y \phi_{j\sigma} = i\int d\tau\,dx\,dy(u^s\partial_\tau\theta^o + u^o\partial_\tau\theta^s)$. Collecting all contributions we find the total action $S = S_\tau + \int d\tau\mathcal{H}_{\text{eff}}$. We integrate out the staggered fields $\theta^s$ and $\phi^s$ and after straightforward algebra we get the following effective action for the stripe phases

$$S = S_u(u^o) + S_\theta(\theta^o) + S_{\text{int}}(u^o, \theta^o)$$

$$S_u(u) = \frac{1}{2}\int_{\vec{r}\tau}\left(\frac{1}{m_2^2}(\partial_\tau u)^2 + K_x(\nabla_x u)^2 + K_y(\nabla_y^2 u)^2\right)$$

$$S_\theta(\theta) = \frac{1}{2}\int_{\vec{r}\tau}\left(\frac{1}{m_1^2}(\partial_\tau \theta)^2 + C_x(\nabla_x \theta)^2 + C_y(\nabla_y \theta)^2 \right.$$

$$\left. - 2t\cos(\theta - Px)\right) \quad (4)$$

$$S_{\text{int}}(u,\theta) = \int_{\vec{r}\tau}\left(D(\nabla_x\theta)^2(\nabla_y u)^2 - \frac{i\Gamma_0}{m_1^2}(\nabla_x u)(\partial_\tau\theta)\right)$$

where $\int_{\vec{r}\tau} = \int dx\,dy\,d\tau$, the last term gives the usual Berry's phase coupling of the $z$ component of magnetization ($\nabla_x u$), and $K_x = K'_x - \Gamma_0^2/m_1^2$. Corrections to $K_x$ due to $\Gamma_0$ can in principle lead to its sign change, which corresponds to a change of the stripe period. For simplicity we assume that $\Gamma_0$ is small everywhere in this paper so $K_x$ remains positive.

For the ISpSt phases IV and V there is average winding of the isospin phase $\theta^o = Px$, with massive fluctuations around it. Stripe positional order has gapless excitations, reflecting broken translational symmetry in the stripe phase. So at $T = 0$ we find the following spectrum $\omega_1(q_x = 0, q_y) = m_1(t + C_y q_y^2)^{1/2}$, $\omega_2(q_x = 0, q_y) = m_2(DP^2)^{1/2}|q_y|$, $\omega_1(q_x, q_y = 0) = m_1(t + (C_x + m_2^2\Gamma_0^2/m_1^4)q_x^2)^{1/2}$, and $\omega_2(q_y = 0, q_x) = m_2 K_x^{1/2}(1 - m_2^2\Gamma_0^2 q_x^2/m_1^4 t)^{1/2}|q_x|$, where $\omega_1(\vec{q})$ and $\omega_2(\vec{q})$ describe isospin and stripe position fluctuations, respectively. At finite temperature, coupling between the stripe positional order and isospin is irrelevant. The former is described by a two-dimensional XY model, which should undergo a Kosterlitz–Thouless transition at finite temperature. At this transition the system establishes quasi-long range order in the stripe positions [9], and a true long-range order may only appear at $T = 0$.

In the ISp phase II we do not have stripes, which can be thought of as melting the long range positional order in $u$ by quantum fluctuations. The description of such a phase may be readily obtained using duality transformation of the $2 + 1$ dimensional XY model for $\tilde{S}_u$ (see Ref. [32] and references therein). Disregarding the anisotropies we find

$$\tilde{S}_{u,\text{dual}} + \tilde{S}_{\text{int,dual}} = \int_{\vec{r}\tau}\left(\frac{\kappa_\mu}{2}|(\partial_\mu - i2\pi a_\mu)\Psi_\nu^u|^2\right.$$

$$\left. \times\,+\,\mathcal{L}_{\text{GL}}(|\Psi_\nu^u|) + \frac{1}{2\kappa_0}f_{\mu\nu}^2 - \frac{i\Gamma_0}{m_1^2}(\nabla_x\theta^o)\epsilon_{\alpha\beta}\nabla_\alpha a_\beta\right). \quad (5)$$

Here $\mu$ and $\nu$ include temporal and spatial components, and $\alpha$ and $\beta$ spatial components only. $\mathcal{L}_{\text{GL}}$ corresponds to the usual Ginzburg–Landau type Lagrangian. The field $\Psi_\nu^u$ creates a dislocation in the stripe lattice and is dual to $e^{i2\pi u^o/a}$. When the crystal is melted by quantum fluctuations $\langle\Psi_\nu^u\rangle \neq 0$, so the gauge field $a_\mu$ is massive by Meissner effect. We can integrate it out and find that it contributes terms to the effective action for $\theta$: $\Delta S(\theta) = -\gamma_1/2(\nabla_x^2\theta)^2 - \gamma_2/2(\nabla_x\nabla_y\theta)^2$. Stripe position fluctuations make isospin fluctuations softer, but only as $q^4$.

In the ISt phase III at any finite $B_\parallel$ there is average winding of the isospin phase $\theta^o$ along $x$ at wavevector $\tilde{Q}$ smaller than $P$ (assuming $\vec{B}_\parallel \| \hat{y}$). Fluctuations of this variable relative to its mean-field configuration $\tilde{\theta} = \theta^o - \tilde{Q}x$ are massless ($t \to 0$) in the long-wavelength limit, and orientation of the stripe order is fixed by finite $\tilde{Q}$. The

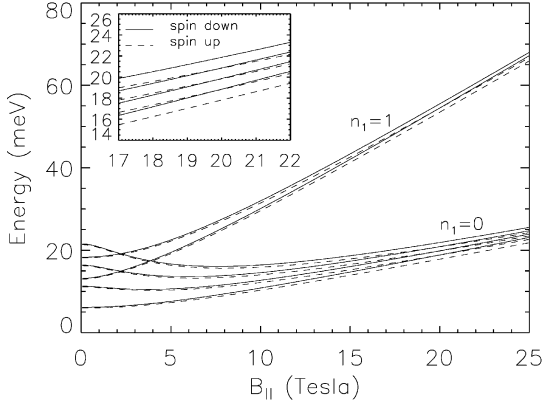

Fig. 6. Noninteracting single electron energy levels in a parabolic confinement potential with a tilted magnetic field as a function of parallel magnetic field, $B_\parallel$. The perpendicular magnetic field is 3 T, and the confinement potential is 7 meV. At even filling factor, the level crossings between electrons of opposite spins occurs at $B_\parallel = 20\,T$.

effective action for this phase is

$$\tilde{S} = \tilde{S}_u(u^o) + \tilde{S}_\theta(\tilde{\theta}) + \tilde{S}_{\text{int}}(u^o, \tilde{\theta})$$

$$\tilde{S}_u(u) = \frac{1}{2} \int_{\vec{r}\tau} \left( \frac{1}{m_2^2}(\partial_\tau u)^2 + K_x(\nabla_x u)^2 + D\tilde{Q}^2(\nabla_y u)^2 + K_y(\nabla_y^2 u)^2 \right)$$

$$\tilde{S}_\theta(\tilde{\theta}) = \frac{1}{2} \int_{\vec{r}\tau} \left( \frac{1}{m_1^2}(\partial_\tau \tilde{\theta})^2 + C_x(\nabla_x \tilde{\theta})^2 + C_y(\nabla_y \tilde{\theta})^2 + 2C_x \tilde{Q}(\nabla_x \tilde{\theta}) \right) \quad (6)$$

$$\tilde{S}_{\text{int}}(u, \tilde{\theta}) = \int_{\vec{r}\tau} \left( D(\nabla_x \tilde{\theta})^2 (\nabla_y u)^2 - \frac{i\Gamma_0}{m_1^2}(\nabla_x u)(\partial_\tau \tilde{\theta}) \right)$$

At zero temperature the system has Goldstone modes coming from two kinds of broken symmetry: spin $XY$ symmetry broken by the expectation value of $\theta$ (this mode has $\omega_1(\vec{q})$) and translational symmetry broken by the stripe positions $u^0$ (this mode has $\omega_2(\vec{q})$). We find $\omega_1(q_x = 0, q_y) = m_1 C_y^{1/2}|q_y|$, $\omega_2(q_x = 0, q_y) = m_2(D\tilde{Q}^2)^{1/2}|q_y|$, $\omega_1(q_x, q_y = 0) = m_1 C_x^{1/2}(1 - \lambda)^{1/2}|q_x|$, $\omega_2(q_x, q_y = 0) = m_2 K_x^{1/2}(1 + \lambda)^{1/2}|q_x|$, where $\lambda = \Gamma_0^2 m_2^2/(m_1^2(K_x m_2^2 - C_x m_1^2)$. In the absence of disorder there can then be two separate Kosterlitz–Thouless type transitions for stripe and isospin orders.

Disorder will have a strong effect on the stripe phases. If impurities interact differently with electrons in the two layers, they will couple to the stripe position $u$. The model $\tilde{S}_u$ with disorder is a problem of melting of a two-dimensional solid on a disordered substrate. The latter was studied by Carpentier and Le Doussal [33], who showed that the melting transition will be replaced by a sharp crossover between a high temperature liquid with thermally induced dislocations, and a low temperature glassy phase with disorder induced dislocations [33]. When the stripes form a glassy configuration this will have an effect on the spin order as well. From the last line of Eq. (4) we can see that when $\nabla_x u$ becomes random it produces a random Berry's phase for $U(1)$ order parameter $\theta^o$. The latter is unimportant in the phase IV, where $\theta^o$ is massive, but may give rise to the Bose glass phase [34,35] in phase III.

## 4. Wide well systems at $\nu = 2N + 2$

Another class of QH systems at integer filling factors, where we can expect the appearance of stripe phases are electrons in a wide well in tilted magnetic field. As an example we consider a parabolic confining potential, which has energy levels in tilted magnetic field (for noninteracting electrons) of the form $E_{n_1,n_2,\sigma_z} = \omega_1(n_1 + 1/2) + \omega_2(n_2 + 1/2) - \omega_{\tilde{z}} \sigma_{\tilde{z}}$ (see Ref. [26,36] and references therein), where $\omega_2$ decreases and Zeeman energy $\omega_{\tilde{z}}$ increases with increasing $B_\parallel$ and constant $B_\perp$ (spin quantization axis $\tilde{z}$ is in the direction of the total magnetic field). At some point there is a level crossing between $(n_1, n_2, \downarrow)$ and $(n_1, n_2 + 1, \uparrow)$ orbitals (see Fig. 6). In the experimentally relevant regime of parameters $n_1 = 0$, so from now on we will only be concerned with this case. For a noninteracting system this level crossing would give rise to a first order phase transition at even filling factors [10] with an abrupt change in the spin polarization. When interactions are taken into account, nontrivial many body states may be stabilized due to hybridization between the orbitals $(0, N, \downarrow)$ and $(0, N + 1, \uparrow)$, where $N = \nu/2 - 1$. The Slater determinant wavefunctions for such states are the same as in Eq. (2) with $c_{k\uparrow}^\dagger$ and $c_{k\downarrow}^\dagger$ creating electrons in states with quantum numbers $(0, N, k, \downarrow)$ and $(0, N + 1, k, \uparrow)$, respectively (we use Landau gauge and assume that parallel magnetic field is along $x$, so electron eigenstates have a well defined momentum $k$ in the $y$ direction). The transverse spin order parameter, resulting from mixing of the $(0, N, \downarrow)$ and $(0, N + 1, \uparrow)$ orbitals, changes sign as a function of $z$ (due to a different $z$ dependence of the two orbitals). Hence, when $w_k$ is constant and $\vec{Q} = 0$, we find a state that may be described as a 'canted antiferromagnetic phase' (CAF). It has constant spin polarization $S_{\tilde{z}}(x, y) = \int dz(\Psi_\uparrow^\dagger(\vec{r})\Psi_\uparrow(\vec{r}) - \Psi_\downarrow^\dagger(\vec{r})\Psi_\downarrow(\vec{r}))/2$ and Neel order parameter $N_+(x, y) = \int dz\, \text{sgn}(z)\Psi_\uparrow^\dagger(\vec{r})\Psi_\downarrow(\vec{r})$. The CAF phase is somewhat analogous to the interlayer coherent phase for bilayer systems; however, in this case the relative phase between the two states is not fixed by tunneling (see also Refs. [37–39] and references therein). When $w_k$ is constant and there is finite $\vec{Q}$, we find the 'spin spiral phase' (SSp), that has constant $S_{\tilde{z}}$ and winding in the transverse Neel order $N_+(\vec{r}_\perp) = N_0\, e^{i\vec{Q}\cdot\vec{r}_\perp}$ ($\vec{r}_\perp = (x,y)$). Unlike the ISp, this SSp phase has gapless fluctuations of the transverse Neel order parameter, reflecting the spin $U(1)$ symmetry of the problem (spin rotations around $\tilde{z}$). One can easily show, that for a fixed $0 < w < \pi$ there is always some $\vec{Q}$ vector, perpendicular to $\vec{B}_\parallel$ (i.e. along $\hat{y}$ in our case), that allows one to construct the SSp phase with energy lower than the CAF phase. The origin of the transverse order parameter winding in bilayers and in wide wells is different. In the former case it comes from tunneling, whereas in the latter case it is due to the changes in the Coulomb matrix elements. When we use single electron wavefunctions in the parabolic potential in tilted magnetic field with $\vec{B}_\parallel \| \hat{x}$ (in Landau gauge), we find that the minimum of dispersion for $N_+(\vec{r}_\perp)$ shifted to finite $\vec{Q} = \pm(0, Q)$. When $w_k$ is periodically modulated with $k$ in wavefunction (2) we get spin stripe phases, which have $\langle S_{\tilde{z}}\rangle$ modulated along $x$. A state with stripes and $\vec{Q} = 0$ we call a 'spin stripe phase' (SSt), and a state with $\vec{Q}\|\hat{y}$ we denote a 'spin skyrmion stripe phase' (SSkSt). The latter has finite spin topological density, and therefore has charge density wave order, in addition to the spin density wave order. There is also a 'spin spiral stripe phase' (SSpSt) phase, that has $\vec{Q}\|\hat{y}$ and stripes along the $x$ direction, i.e. parallel to $B_\parallel$. The wavefunction for the SSpSt phase, however, cannot be written in the form of Eq. (1). Finally, the degeneracy of the states $\pm\vec{Q}$ allows a collinear SDW phase with $N_{x,y}(\vec{r}_\perp) \propto \cos(\vec{Q}\cdot\vec{r}_\perp)$, but we will not discuss it in this article.

Within HF variational calculations at $T = 0$ and without taking into account screening by the lower Landau levels we find [26] that in a parabolic confining potential all the nontrivial many-body states are higher in energy than simple polarized incompressible QH states (the latter have noninteracting Landau levels either completely empty or occupied). Taken literally, this would imply that interactions do not alter the scenario where there is a first order transition with a jump in the polarization. Our results indicate, however, that in the vicinity of the HF level crossing, the energy difference between the polarized state and some of the many-body states that we consider (the SSp, the SSpSt, and the SSkSt phases) is small. It is then possible, that when we use a different confining potential, include screening by the lower levels, and/or go beyond mean field HF approximation some of these many body states may very well become lower in energy than the polarized phases.

We find (for transition at $N = 2$, which corresponds to $\nu = 6$) that stripe phases with winding (the SSpSt and the SSkSt phases) are more favorable than the SSt phase with no winding (all the comparisons were done for configurations with equal spin polarization). We find that the SSpSt phase is usually more favorable than the SSkSt phase, although the difference between the two is extremely small. We also find that the SSp phase, that has winding and no stripes, is slightly more favorable than any of the stripe phases. These results will be published elsewhere [26].

The effective action for the stripe phases in a wide well and tilted magnetic field (for both, SSkSt and SSpSt phases) is similar to the action for bilayer first phase and is given by Eq. (6). Variable $u^o$ describes stripe position fluctuations, and $\tilde{\theta}$ corresponds to the fluctuation in the direction of the transverse Neel order parameter relative to its mean-field configuration (the latter has uniform winding along the stripes, as shown on Fig. 1(c)). Both variables describe modes that are soft in the long wavelength limit, with dispersion similar to what has been discussed before for phase III of the bilayer system.

The lowest energy elementary charge excitations in the stripe phases (both spin and isospin) at integer filling factors are solitons of $\theta^o$ on individual domain walls (see, e.g. [9,40]). The easy direction for charge transport (i.e. direction of smaller longitudinal resistance) is along domain walls, so it is along the parallel magnetic field in the spiral stripe phase and perpendicular to it in the skyrmion stripe phase. This observation motivates us to suggest, that the resistance anisotropy observed in Ref. [10], may be due to the formation of a spin skyrmion stripe phase.

It is useful to compare our picture of the SSkSt phase with the domain wall picture suggested by Jungwirth et al. [41] to explain the experiments of Ref. [10]. According to Ref. [41], no nontrivial stripe phases are stabilized around the level crossing of $(0, N, \downarrow)$ and $(0, N + 1, \uparrow)$ orbitals, but disorder gives rise to an appreciable number of domain walls between regions of different spin polarization. Low energy collective excitations of such domain walls are then responsible for the transport anisotropy. Our SSkSt phase can be thought of as a collection of polarized domains (stabilized by Coulomb interaction, rather than disorder), with nontrivial spin winding on the domain walls and low energy excitations in the form of $\tilde{\theta}$ solitons (we note that the textured edges of the QH systems have been discussed in Refs. [42–44]). Hence, although there is different origin of the domain walls in the two scenarios, there may be certain phenomenological similarity in the description of the transport properties. Ref. [41] finds domain walls that are favored in the direction along $B_\parallel$, which implies that the easy direction for transport is along $B_\parallel$. This is consistent with our HF calculations, that show SSpSt phase to be slightly lower in energy than the SSkSt phase. However, the skyrmion stripe phase is more consistent with the resistance anisotropy observed in Ref. [10], therefore we think it is more likely to have been realized in experimental systems. We should also mention that our theory may be applicable to the magnetoresistance anisotropy recently observed [45] in Si/SiGe two-dimensional electron systems in the presence of tilted magnetic fields, but the complication of valley degeneracy in Si makes a straightforward comparison between our theory and Ref. [45] difficult.

Before concluding we point out that our analysis of this section should, in principle, also apply to a system with an odd $\nu$ if the Zeeman coupling is large. For example, for $\nu = 5$ one could imagine a situation where the $n_2 = 3$ Landau orbital level with spin up is degenerate with an $n_2 = 1$ level of spin down (it happens outside the range of Fig. 6). Then we would have a 'base state' where the second and the third Landau levels have filled spin up states, and there is one additional electron per flux quantum divided between the two degenerate states. Our results of this section would work perfectly well for this case although such a situation may be more difficult to achieve experimentally. This and other such level crossing situations will be discussed in details in a forthcoming publication [26].

## 5. Summary

To summarize, we use a HF analysis to discuss the phase diagram of bilayer QH systems at $\nu = 4N + 1$, establish the possibility of several distinct ISts, and derive an effective theory for them. We point out that similar stripe phases may exist in wide well systems at $\nu = 2N + 2$ and propose that the resistance anisotropy recently observed in Ref. [10] may be due to the formation of a spin skyrmion stripe phase.

## Acknowledgments

This work is supported by the NSF grant DMR 99-81283, US-ONR, DARPA, ARDA, and the Harvard Society of Fellows. We acknowledge useful discussions with C. Kallin, S. Kivelson, A. Lopatnikova, A. MacDonald, I. Martin, C. Nayak, L. Radzihovsky, and S. Simon.

## References

[1] D.C. Tsui, H.L. Stormer, A.C. Gossard, Phys. Rev. Lett. 48 (1982) 1559.
[2] M.P. Lilly, K.B. Cooper, J.P. Eisenstein, L.N. Pfeiffer, K.W. West, Phys. Rev. Lett. 82 (1999) 394.
[3] R.R. Du, D.C. Tsui, H.L. Stormer, L.N. Pfeiffer, K.W. Baldwin, K.W. West, Solid State Commun. 109 (1999) 389.
[4] M.P. Lilly, K.B. Cooper, J.P. Eisenstein, L.N. Pfeiffer, K.W. West, Phys. Rev. Lett. 83 (1999) 824.
[5] M.M. Fogler, A.A. Koulakov, B.I. Shklovskii, Phys. Rev. B 54 (1996) 1853.
[6] R. Moessner, J.T. Chalker, Phys. Rev. B 54 (1996) 5006.
[7] T. Jungwirth, A.H. MacDonald, L. Smrcka, S.M. Girvin, Physica E 6 (2000) 43.
[8] A.H. MacDonald, M.P.A. Fisher, Phys. Rev. B 61 (2000) 5724.
[9] L. Brey, H.A. Fertig, Phys. Rev. B 62 (2000) 10268.
[10] W. Pan, H.L. Stormer, D.C. Tsui, L.N. Pfeiffer, K.W. Baldwin, K.W. West, Phys. Rev. B 64 (2001) 121305.
[11] J. Eisenstein, in: S. Das Sarma, A. Pinczuk (Eds.), Perspectives in Quantum Hall Effect, Wiley, New York, 1997.
[12] S. Girvin, A.H. MacDonald, in: S. Das Sarma, A. Pinczuk (Eds.), Perspectives in Quantum Hall Effect, Wiley, New York, 1997.
[13] C.B. Hanna, A.H. MacDonald, S.M. Girvin, Phys. Rev. B 63 (2001) 125305.
[14] X.G. Wen, A. Zee, cond-mat/0110007.
[15] E. Demler, C. Nayak, S. Das Sarma, Phys. Rev. Lett. 86 (2001) 1853.
[16] Y.B. Kim, C. Nayak, E. Demler, N. Read, S. Das Sarma, Phys. Rev. B 63 (2001) 205315.
[17] A. Stern, S. Das Sarma, M.P.A. Fisher, S.M. Girvin, Phys. Rev. Lett. 84 (2000) 139.
[18] B.I. Halperin, Helvetica Physica Acta 56 (1983) 75.
[19] M. Rasolt, B.I. Halperin, D. Vanderbilt, Phys. Rev. Lett. 57 (1986) 126.
[20] H.A. Fertig, Phys. Rev. B 40 (1989) 1087.
[21] I.B. Spielman, J.P. Eisenstein, L.N. Pfeiffer, K.W. West, Phys. Rev. Lett. 84 (2000) 5808.
[22] N. Read, Phys. Rev. B 52 (1995) 1926.
[23] S.L. Sondhi, A. Karlhede, S.A. Kivelson, E.H. Rezayi, Phys. Rev. B 47 (1992) 16419.
[24] L. Radzihovsky, Phys. Rev. Lett. 87 (2001) 236802.
[25] R. Abolfath, L. Radzihovsky, A. MacDonald, cond-mat/0110049.
[26] D.W. Wang, E. Demler, S. Das Sarma, B.I. Halperin, in preparation.
[27] C.L. Kane, M.P.A. Fisher, in: S. Das Sarma, A. Pinczuk (Eds.), Perspectives in Quantum Hall Effect, Wiley, New York, 1997.
[28] A. Lopatnikova, S.H. Simon, B.I. Halperin, X.G. Wen, Phys. Rev. B 64 (2001) 155301.
[29] R. Cote, H. Fertig, Phys. Rev. B 65 (2002) 085321.
[30] D.G. Barci, E. Fradkin, S.A. Kivelson, V. Oganesyan, cond-mat/0105448.
[31] M. Fogler, cond-mat/0107306, cond-mat/0111001, cond-mat/0111049.
[32] L. Balents, M.P.A. Fisher, C. Nayak, Int. J. Mod. Phys. B 12 (1998) 1033.
[33] D. Carpentier, P. Le Doussal, Phys. Rev. Lett. 81 (1998) 1881.
[34] M.P.A. Fisher, P.B. Weichman, G. Grinstein, D.S. Fisher, Phys. Rev. B 40 (1989) 546.
[35] E. Demler, S. Das Sarma, Phys. Rev. Lett. 83 (1999) 168.
[36] M.P. Stopa, S. Das Sarma, Phys. Rev. B 47 (1993) 2122.
[37] S. Das Sarma, E. Demler, Solid State Commun. 117 (2001) 141.
[38] L. Zheng, R.J. Radtke, S. Das Sarma, Phys. Rev. Lett. 78 (1997) 2453.
[39] S. Das Sarma, S. Sachdev, L. Zhang, Phys. Rev. B 58 (1998) 4672.
[40] V.I. Falko, S.V. Iordansky, Phys. Rev. Lett. 82 (1999) 402.
[41] T. Jungwirth, A.H. MacDonald, E.H. Rezayi, Physica E 12 (2002) 1.
[42] A. Karlhede, K. Lejnell, S.A. Kivelson, S.L. Sondhi, Phys. Rev. Lett. 77 (1996) 2061.
[43] A. Karlhede, K. Lejnell, S.L. Sondhi, Phys. Rev. B 59 (1999) 10183.
[44] A. Karlhede, K. Lejnell, S.L. Sondhi, Phys. Rev. B 60 (1999) 15948.
[45] U. Zeitler, H.W. Schumacher, A.G.M. Jansen, R.J. Haug, Phys. Rev. Lett. 86 (2001) 866.